\documentclass[useAMS,usenatbib]{mn2e}
\input psfig.sty

\usepackage[english]{babel}
\usepackage{aas_macros}
\usepackage{graphicx}
\usepackage{hyperref}
\usepackage{amssymb}
\usepackage{amsmath}
\usepackage{amsfonts}
\newcommand{\kg}{\kappa_{\rm g}}
\newcommand{\kng}{\kappa_{\rm ng}}
\newcommand{\kln}{\kappa_{\rm ln}}

\title[Information in Weak Lensing: Wavelet Method]
{Information Content in the Angular Power Spectrum of Weak Lensing:
Wavelet Method}

\author[Hao-Ran Yu et al.]
{Hao-Ran Yu$^{1,2}$, Joachim Harnois-D\'{e}raps$^{2,3}$, Tong-Jie
Zhang\thanks{E-mail: tjzhang@bnu.edu.cn}$^{1,4}$ and Ue-Li
Pen$^{2}$\\
$^1$Department of Astronomy, Beijing Normal University, Beijing, 100875, P. R. China; tjzhang@bnu.edu.cn\\
$^2$Canadian Institute for Theoretical Astrophysics, University of
Toronto, M5S 3H8, Ontario, Canada\\
$^3$Department of Physics, University of Toronto, M5S 1A7, Ontario,Canada\\
$^4$Center for High Energy Physics, Peking University, Beijing, 100871, P.R. China\\
}

\begin{document}

\date{}

\maketitle

\label{firstpage}

\begin{abstract}
We quantify the performance of a non-linear Wiener filter,
constructed in wavelet space, at recovering some of the Fisher
information that was lost in the weak lensing convergence field. The
proposed method consists in a separation of the original field into
the sum of a Gaussian and a non-Gaussian contribution. After
filtering an ensemble of such fields, which are obtained from
$N$-body simulations, we find that we can recapture about four times
more Fisher information, an effect that can potentially improve by a
significant amount the constraining  power of weak lensing surveys
on cosmological parameters, including the dark energy equation of
state  $\omega$. We compare this performance with that of a
logarithmic mapping and find that the wavelet method can recover up
to three times more information.
\end{abstract}


\section{Introduction}\label{s.intro}
Mapping the mass distribution of matter in the universe has been a
major challenge and focus of modern observational cosmology
\citep{1993ApJ...404..441K,1999A&A...345...17B,1999toc..conf..211M,
2001PhR...340..291B,2003ARA&A..41..645R,
2010RPPh...73h6901M,2010GReGr..42.2177H}. Because the light to mass
bias is rather uncertain, the only direct procedure to weigh the
universe is by measuring the deflection of light caused by the
presence of matter between the source and the observer.
In particular, the statistics of gravitational lensing can serve as
a powerful probe of the mass distribution of the universe
\citep{2001A&A...374..757V,2010APh....32..340V,2007MNRAS.381..702B}.

It was recently realized that weak lensing could also provide
insight on dark energy via the measurement of the growth function
\citep{2002PhRvD..65f3001H,2006astro.ph..9591A,2010GReGr..42.2177H},
and an international effort was put into motion in order to measure
weak lensing signal with unprecedent accuracy and resolution
\citep{2009arXiv0912.0201L,2010ASPC..430..266B,2010arXiv1008.4936G}.
Alternatives techniques such as the redshift distance measurements
of supernovae \citep{1999ApJ...517..565P}, the detection of baryonic
acoustic oscillations (BAO) in galaxy surveys
\citep{2007ApJ...657...51P, Eisenstein:2005su}, the measurement of
the growth factor from clusters \citep{2005RvMP...77..207V} and the
detection of weak lensing signal \citep{2006ApJ...647..116H} have
already set tight constraints on the dark energy equation of state
$\omega$ \citep{2008AJ....135..512O}, and the weak lensing
contribution provides a complimentary approach as it is sensitive to
different systematic uncertainties. The goal of the combined dark
energy experiments is to maximize a collective ``figure-of-merit'',
which influenced he design of most of the future experiments like
LSST\footnote{http://www.lsst.org/lsst/}
\citep{2009arXiv0912.0201L},
EUCLID\footnote{http://www.congrex.nl/09c08/}
\citep{2010ASPC..430..266B},
JDEM\footnote{http://science.nasa.gov/missions/jdem/}
\citep{2010arXiv1008.4936G},
CHIME\footnote{http://www.physics.ubc.ca/chime/}
\citep{2006astro.ph..6104P},
SKA\footnote{http://www.skatelescope.org/}
\citep{2007HiA....14..539S,2009IEEEP..97.1482D},
BOSS\footnote{http://cosmology.lbl.gov/BOSS/}
\citep{2009astro2010S.314S} and PAN
STARRS\footnote{http://pan-starrs.ifa.hawaii.edu/public/}.


It was soon realized in the BAO community that the constraining
strength of these surveys depends directly on the amount of Fisher
information \citep{Fisher1935,1997ApJ...480...22T}, i.e.
statistically independent Fourier modes contained in the
measurements of the power spectrum bands, and one needs to maximize
that information in order to minimize the uncertainty on
cosmological parameters. Counting these Fourier modes is straight
forward when the underlying density field is Gaussian, however
departures from Gaussianity are to be expected,  as arising from the
non-linear gravitational collapse of the density field
\citep{2002PhRvD..65f3001H}. In the theory of structure formation,
large scales structures grow from an initially linear Gaussian
random field, which progressively becomes non-linear through
gravitational instabilities, starting from the smallest scales. Only
the largest scales of the field remains intrinsically Gaussian,
while the non-linear Fourier modes start to couple together
\citep{2003ApJ...598..818Z}.

\cite{2005MNRAS.360L..82R,2006MNRAS.371.1205R} first measured the
amount of Fisher information about the power spectrum amplitude
contained in the matter field, as a function of scale, from an
ensemble of 400 $N$-body simulations. They found that in the largest
scales, the Fisher information grows in a manner consistent with a
Gaussian random field. However, they observed departures from
Gaussianity in the trans-linear regime, in the form of an
information plateau, followed by a second rise on much smaller
scales. This was later interpreted in terms of the halo model as a
transition between the two-halo and the one-halo terms
\citep{2006MNRAS.370L..66N,2007MNRAS.375L..51N}.

This loss of Fisher information is an undesired effect, in the sense
that it is equivalent to a reduction of the survey effective volume,
and many strategies have been proposed to recover some of the erased
cosmic information. \cite{1992MNRAS.254..315W} used a method called
Gaussianization, which is a monotonic transformation of the smoothed
galaxy distribution that reconstructs primordial density
fluctuations. Running $N$-body simulations backwards in time
\citep{2000ASPC..201..282G}, or density field reconstruction from
linear theory
\citep{2007ApJ...664..675E,2009PhRvD..80l3501N,2009PhRvD..79f3523P,
2011arXiv1106.5548N} has also been successful at recovering parts of
the lost information. More recently, it was found that a logarithmic
transformation of density fields appears to be effective on
trans-linear scales \citep{2009ApJ...698L..90N}.
\cite{2011ApJ...731..116N} have successfully recovered some of the
Fisher information with a Gaussianization method that takes the
Poisson noise into account. Non-linear Wiener filters, which can be
designed to decompose a density field into Gaussian and non-Gaussian
parts, are also among the best techniques found so far
\citep{2011ApJ...728...35Z}. All of these techniques somehow attempt
to diagonalize the covariance matrix of the matter power spectra by
bringing back to the 2-point function some of the information that
had leaked to higher order terms.

The next step was to measure the impact of these non-Gaussianities
on cosmological parameters. It was first shown that their inclusion
has only a minor impact on the constraining power about the BAO
dilation scale, when both the power spectrum and the covariance
matrix are obtained from $N$-body simulations
\citep{2011ApJ...726....7T}. In current data analyses, however, the
power spectrum is often estimated with techniques that assume
Gaussianity in the matter field
\citep{1994ApJ...426...23F,1996ApJ...465...34V}, which have the
unfortunate effect of producing sub-optimal estimates of the mean
\citep{2006PhRvD..74l3507T}. In the case where the survey selection
function is complex, the estimate is likely to be biased
\citep{2011arXiv1109.5746H}. In regards with the sub-optimal
measurement, it was recently shown that the error bars on the BAO
dilation scale that are {\it consistent} with a sub-optimal
measurement of the power spectrum might be significantly larger,
compared to those obtained under standard Gaussian prescription
\citep{2011arXiv1106.5548N}. In the era of precision cosmology,
these few percent level effects need to be considered when
constraining dark energy  parameters.

In the pursuit of robustness and accuracy in weak lensing analyses,
equal considerations must be granted to non-Gaussianities. Shear and
convergence maps are indeed sensitive to the non-linear regime,
typically at low and intermediate distances. For instance, it was
recently shown that Fisher information -- about the amplitude of the
lensing power spectrum -- contained in convergence fields was also
departing from the information of a Gaussian field
\citep{2009arXiv0905.0501D,2010PhRvD..81l3015L}. It was furthermore,
shown that these deviations could potentially impact the
constraining power on the dark energy equation of state.
The covariance matrix in the weak lensing angular power spectrum
indeed shows strong correlations across different scales, as
confirmed by \citep{2011ApJ...729L..11S}.

Since then, much efforts have been made to Gaussianize the lensing
fields as well, in an attempt to recover some information and thus
improve the precision on current and future measurements of
cosmological parameters. \citep{2011ApJ...729L..11S} have also shown
that the method of logarithmic mapping is able to pump back some of
the information lost. Such transformation was shown to suppress the
non-Gaussian contribution in the statistics of higher order
cumulants, and can also suppress the bispectrum
\citep{2011arXiv1103.2858Y}. \cite{2011arXiv1104.1399J} also showed
that a Cox-Box transformation can also Gaussianize the fields and
restore about the same amount of information lost in lensing fields,
compared to logarithmic mapping. Those techniques basically
reconstruct a probability distribution function (PDF) of the
$\kappa$ field which is much closer to that of a Gaussian
\citep{2011ApJ...729L..11S,2011arXiv1104.1399J}.

In the effort to Gaussianize the fields, non-linear Wiener filters
are also a promising technique and work especially well with wavelet
transforms which are well suited for extracting multiscale
information \citep{1998World.Sci..2A}. Moreover, they were found to
offer better performances over the standard Fourier basis if the
data are intermittent in nature \citep{1999RSPTA.357.2561P}. They
are proved to be successful in recovering Fisher information about
the amplitude of the power spectrum of the dark matter field
\citep{2011ApJ...728...35Z}, and could potentially outperform other
Gaussianization techniques of weak lensing maps. In this paper, we
thus construct a non-linear filtering method similar to that of
\cite{2011ApJ...728...35Z} to Gaussianize the $\kappa$ fields, and
we compare the increase the Fisher information contained in
simulated maps with other methods.

\begin{figure*}
    \centerline{
    \psfig{figure=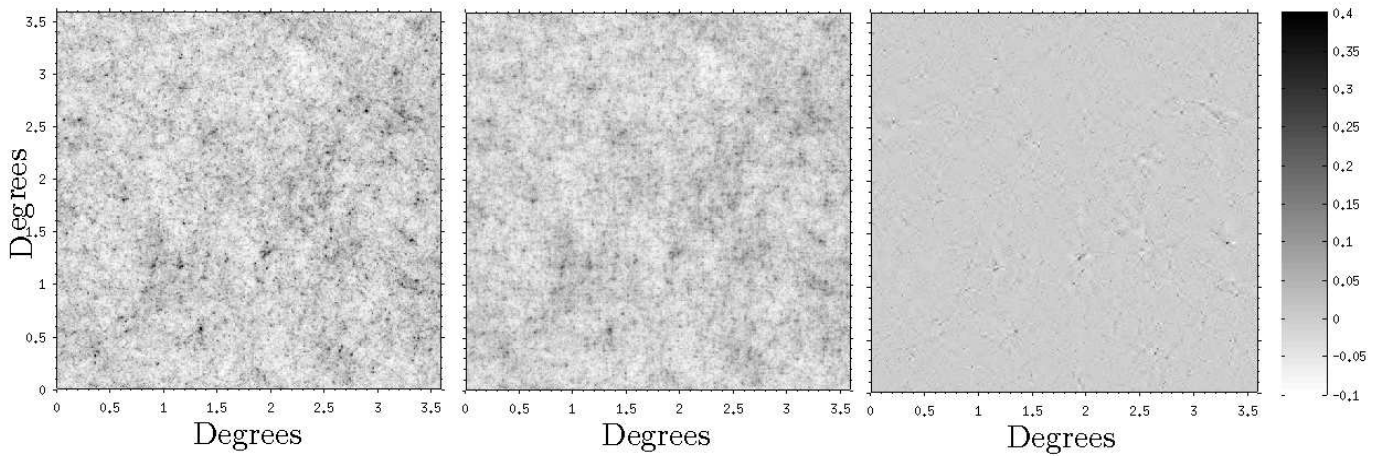,width=8.0truein,height=2.8truein}
    \hskip 0.0in}
    \caption{Random selected $\kappa$ field constructed from $N$-body simulation
    (left panel).
    After the non-linear wavelet Wiener filtering, this field is decomposed
    into $\kg$ (middle panel) and $\kng$ (right panel).}
    \label{f.maps}
\end{figure*}

The outline of the paper is as follows: we first describe our
strategy to construct weak lensing maps from $N$-body simulations in
\S \ref{s.simulation}. We discuss in \S \ref{s.filtering} the
application of the discrete wavelet transform (DWT) and Bayesian
theory to filter out the non-Gaussian component of these fields. In
\S \ref{s.analysis}, we calculate the information content about the
power spectra in the Gaussianized $\kappa$ fields. Discussion and
conclusion are presented in \S \ref{s.conclusion}.


\section{Simulations and construction of convergence maps}\label{s.simulation}
We ran a series of 185 $N$-body simulations to generate convergence
maps, using the fast parallel {\tt cubep3m}
\citep{2005NewA...10..393M}, a particle-mesh Poisson solver that has
sub-grid resolution, thanks to the calculation of the exact Newton
force between particles. It was run with $1024^3$ particles on eight
IBM nodes of the Tightly Coupled System on SciNet, each node being
equipped with 32 cores, 128Gb of RAM, with an infiniband connection
across nodes \citep{Scinet}. We output a series of periodic surface
density on $2048^2$ pixel maps, along the three orthogonal
directions of the cube at each specified redshifts.

The simulations were designed to optimize the usage of the simulated
comoving volume in the construction of the past light cone,  which
consist of a juxtaposition of cubes of $231.1 \mbox{Mpc}/h$ per side
for $z>1$, and of $147.0 \mbox{Mpc}/h$ per side for lower redshifts.
The light cone opening  angle was set to 3.58 degrees, which exactly
touches the edges of the small box at $z=1$, then propagates into
the larger boxes until $z=2$, beyond which we used the periodicity
of the simulations to populate the volumes. This creates repeated
structures, which increase the correlation across different Fourier
modes in the mass density at the percent level. But this has
negligible impact on our results since the lensing kernel strongly
suppresses the contributions from such high redshifts. At most, such
increased correlation indeed accentuates the non-Gaussian features,
but this effect gets strongly suppressed by the projection along the
line of sight.

Simulations started at an initial redshift $z_{\rm i}=40$ for the
lower redshift boxes and $200$ otherwise. The cosmological
parameters used are $\Omega_{\rm M}=0.279$, $\Omega_\Lambda=0.721$,
with Hubble constant $h=0.701$, and we obtained a transfer function
from {\tt CAMB} with $n_{s}=0.96$. The power spectrum normalization
was then specified to be $\sigma_8=0.815$.

The convergence $\kappa$ field is obtained from the projection of
the matter overdensity $\delta$ along the line of sight $\theta$,
weighted by the lensing geometry and, potentially, a source-galaxy
distribution. It can be expressed as
\begin{equation}
    \kappa(\theta,\chi)=\int_0^{\chi} W(\chi')
    \delta(\chi',r(\chi')\theta){\rm d}\chi',
\end{equation}
where $\chi$ is the comoving distance in unit of $c/H_0$, and
$H_0=100\  h \ {\rm km\ s}^{-1}\ {\rm Mpc}^{-1}$. The weight
function $W(\chi)$ is\
\begin{equation}
W(\chi)=\frac{3}{2}\Omega_{\rm M}g(\chi)(1+z)
\end{equation}
where $g(\chi)$ is determined by  the source galaxy distribution
function $n(z)$ and the lensing geometry:
\begin{equation}
g(\chi)=r(\chi) \int_{\chi}^{\infty}
n(\chi')\frac{r(\chi'-\chi)}{r(\chi')}{\rm d}\chi'. \label{eqn:gx}
\end{equation}
Here, $r(\chi)$ is the radial coordinate and is equal to $\chi$ for
the flat geometry we consider.

The density fields are converted into convergence maps by stacking
the images with the appropriate geometrical weights $W(z)$, through
the comoving volume contained in the past light cone. This tiling
method was developed in \cite{1998ApJ...506...64S} and  assumes both
the thin lens and Born approximations. At each lens plane, we choose
randomly one of the $x$-, $y$- and $z$- directions, and we shift the
centre of the plane. This effectively suppresses much of the
correlation that exists across the lenses. We then interpolate the
lens onto a  pixel map of constant angular size. Since we chose a
uniform galaxy distribution, and a source placed at $z_{s} = 3.0$,
most of the lensing contribution comes from  $z\sim 1.5$. One of
these $\kappa$ map is shown in the left panel of Fig.\ref{f.maps}.

\section{Wavelet Non-Linear Wiener Filtering of $\kappa$ fields}\label{s.filtering}

In this section, we first briefly review the properties of the
wavelet transform that are relevant to our discussion, then we
describe our non-linear Wiener filter, and finally we quickly
present the log transform, which has been previously used to
Gaussianize the $\kappa$ fields and against which we compare our
results in section \ref{s.analysis}.

\subsection{Discrete Wavelet Transform (DWT)}

Similar to the Fast Fourier Transform (FFT), the Discrete Wavelet
Transform (DWT) is an invertible, linear operation, and can be
considered as a rotation in function space.
We consider in this paper the Daubechies-4 (hereafter DB-4) wavelet
basis \citep{Daubechies1992}, which contains families of {\it
scaling} functions $\phi$ and {\it difference} functions (or {\it
wavelet} functions) $\psi$ that are orthogonal, continuous and have
compact support. In a DWT decomposition, the data which, in our
case, have $2^J$ grid elements per dimension, are thus expanded into
a combinations of these orthogonal basis, and weighted by wavelet
coefficients $\epsilon$.

In two dimensions, each wavelet coefficient thus depend on two scale
indices $(j_1,j_2)$ -- controlling the {\it dilation} of the wavelet
DB-4 functions -- and two location indices $(l_1,l_2)$  --
controlling its translation. On a given dimension, the grid scale
corresponding to a specified dilation is  $L/2^j$, where $L=2048$ in
our case. Similarly, the index $l$ corresponds to the spatial
location comprised in the range $lL/2^j<x<(l+1)/2^j$. The
2-dimensional convergence field $\kappa(\vec{\theta}(x_1,x_2))$
 can be expanded as
\begin{eqnarray}\label{e.expansion}
    \nonumber
    \vec{\kappa}(x_1,x_2)=
    \sum_{l_1=0}^{1}\sum_{l_2=0}^{1}\epsilon_{0,0;l_1,l_2}\phi_{0,l_1}(x_1)\phi_{0,l_2}(x_2) \\
    \nonumber
    +\sum_{j_1=0}^{J-1}\sum_{l_1=0}^{2^{j_1-1}}\sum_{l_2=0}^{1}
    \tilde{\epsilon}^{(1)}_{j_1,0;l_1,l_2}\psi_{j_1,l_1}(x_1)\phi_{0,l_2}(x_2) \\
    \nonumber
    +\sum_{l_1=0}^{1}\sum_{j_2=0}^{J-1}\sum_{l_2=0}^{2^{j_2-1}}
    \tilde{\epsilon}^{(2)}_{0,j_2;l_1,l_2}\phi_{0,l_1}(x_1)\psi_{j_2,l_2}(x_2) \\
    +\sum_{j_1=0}^{J-1}\sum_{l_1=0}^{2^{j_1-1}}\sum_{j_2=0}^{J-1}\sum_{l_2=0}^{2^{j_2-1}}
    \tilde{\epsilon}_{j_1,j_2;l_1,l_2}\psi_{j_1,l_1}(x_1)\psi_{j_2,l_2}(x_2),
\end{eqnarray}
where
\begin{equation}\label{}
    \phi_{j,l}(x)=\sqrt{\frac{2^j}{L}} \phi(2^jx/L-l)
\end{equation}
\begin{equation}
    \psi_{j,l}(x)=\sqrt{\frac{2^j}{L}} \psi(2^jx/L-l).
\end{equation}
With the combination of these two basis functions, scaling function
coefficients (hereafter SFC's) $\epsilon$ 's and three kinds of
wavelet function coefficients (hereafter WFC's) $\tilde{\epsilon}$
's can be calculated by
\begin{equation}\label{e.SFCs}
    \epsilon_{j_1,j_2;l_1,l_2}=
    \iint \vec{\kappa}(x_1,x_2)\phi_{j_1,l_1}(x_1)\phi_{j_2,l_2}(x_2)
    {\rm d}x_1{\rm d}x_2,
\end{equation}
\begin{equation}\label{e.WFCs(1)}
    \tilde\epsilon^{(1)}_{j_1,j_2;l_1,l_2}=
    \iint \vec{\kappa}(x_1,x_2)\psi_{j_1,l_1}(x_1)\phi_{j_2,l_2}(x_2)
    {\rm d}x_1{\rm d}x_2,
\end{equation}
\begin{equation}\label{e.WFCs(2)}
    \tilde\epsilon^{(2)}_{j_1,j_2;l_1,l_2}=
    \iint \vec{\kappa}(x_1,x_2)\phi_{j_1,l_1}(x_1)\psi_{j_2,l_2}(x_2)
    {\rm d}x_1{\rm d}x_2,
\end{equation}
and
\begin{equation}\label{e.WFCs}
    \tilde\epsilon_{j_1,j_2;l_1,l_2}=
    \iint \vec{\kappa}(x_1,x_2)\psi_{j_1,l_1}(x_1)\psi_{j_2,l_2}(x_2)
    {\rm d}x_1{\rm d}x_2.
\end{equation}

For each simulation, the $\kappa$ field is thus wavelet transformed,
and each of the four kinds of coefficients found in
Eqs.(\ref{e.SFCs}-\ref{e.WFCs}) are stored in a 2-dimensional field,
preserving the grid resolution (see \cite{1998World.Sci..2A,nr} for
more details).

\subsection{Non-linear Wiener Filtering}

Our strategy to construct a non-linear Wiener filter relies on the
fact that in wavelet basis, the non-Gaussianities are clearly
characterized in the PDF of the WFCs
$\tilde\epsilon_{j_1,j_2;l_1,l_2}$, which we obtained from
Eq.(\ref{e.WFCs}). We thus construct our filter by splitting the
wavelet transform of the original map, which we label $K$, into two
components: a Gaussian ($G$) and a non-Gaussian ($N$) map. Namely,
in wavelet space, we have
\begin{equation}\label{dabw}
    K=G+N.
\end{equation}
Since wavelet transforms are linear operations, we can inverse
wavelet transform the above equation and write, in real space,
\begin{equation}\label{dab}
    \kappa=\kappa_{\rm g}+\kappa_{\rm ng}
\end{equation}
where the original map ($\kappa$) is expressed as the sum over a
Gaussian contribution (hereafter $\kappa_{\rm g}$) and a
non-Gaussianized contribution (hereafter $\kappa_{\rm ng}$). Our
goal is thus to design a filter that concentrates most of the
collapsed structure in $\kappa_{\rm ng}$, and thus produces
$\kappa_{\rm g}$ that are closer to linear theory. We perform this
operation on our simulated maps, compute their power spectrum,
construct a covariance matrix and measure the  Fisher information of
both components separately. Then we can finally compare our results
with the unfiltered  maps.

\begin{figure*}
    \centerline{
    \psfig{figure=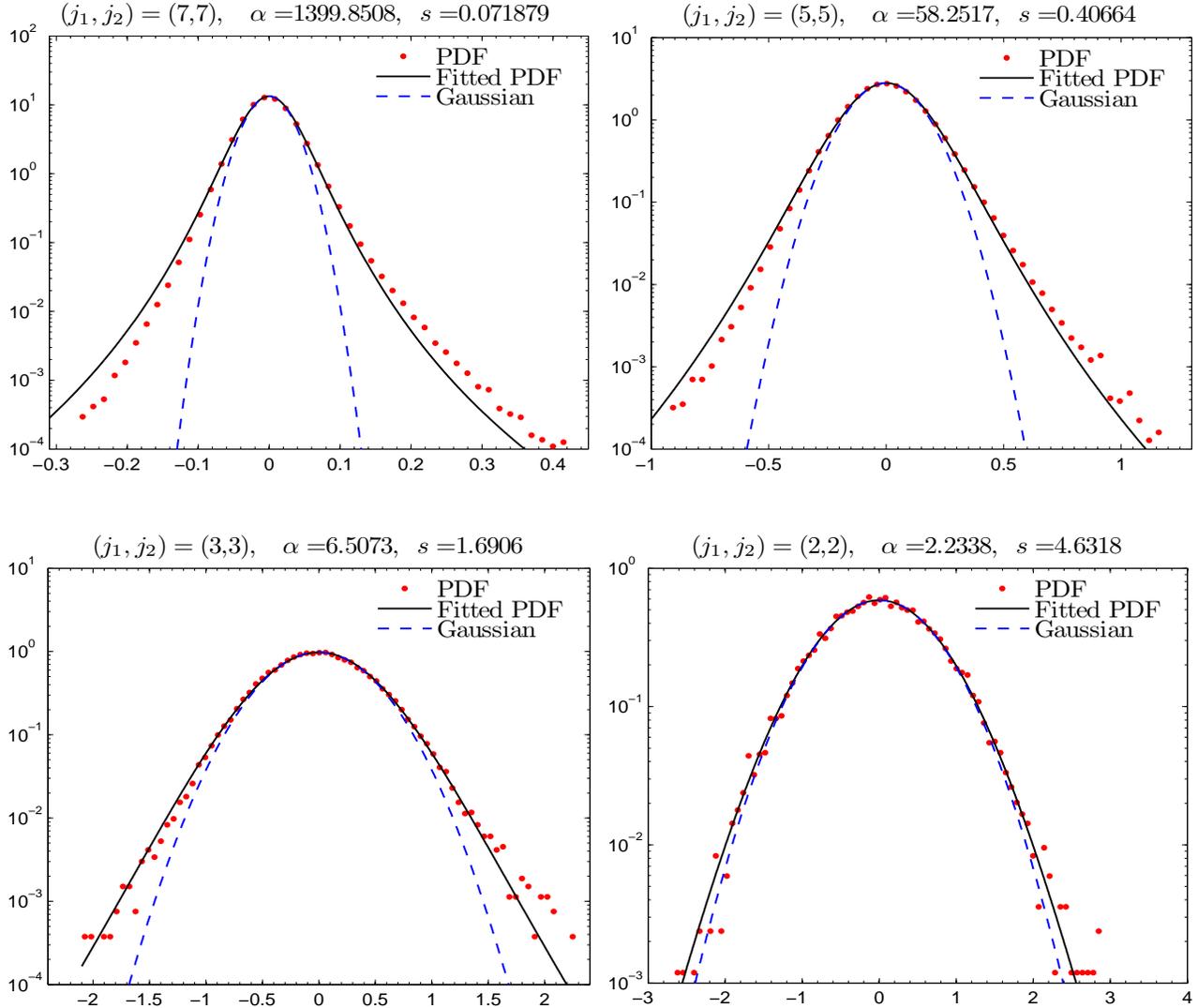,width=8.6truein,height=6.0truein}
    \hskip 0.0in}
    \caption{WFC PDFs (red dots) and their analytical fits by Eq.(\ref{fPDF})
    (black solid lines). Their corresponding Gaussian PDFs, which have the same standard
    deviations with the central regions of the fitted PDFs, are also shown in blue dashed lines.
    The upper left, upper right, bottom left and bottom right panels
    show four different wavelet modes, from smaller (more non-Gaussian)
    to larger scales (more Gaussian), as indicated by the value of
    $(j_1,j_2)$'s in the titles. They are roughly corresponding to multipole
    moment $\ell\simeq9.1\times 10^3$, $2.3\times 10^3$, $5.7\times 10^2$
    and $2.8\times 10^2$, respectively. The two parameters
    $\alpha$ and $s$ for Eq.(\ref{fPDF}) are also shown in the title of each panel.
    }
    \label{f.pdfs}
\end{figure*}

The algorithm we use to devise the filter is a 2-dimensional version
of that presented in \citep{2011ApJ...728...35Z}, which we briefly
describe here again for completeness. The filter acts on each {\it
wavelet mode}, which is defined as a collection of all WFCs having
the same two scale indices $(j_1,j_2)$. Let us focus on a given
wavelet mode, or a pair of scale components, or dilations
$(j_1,j_2)$, whose value at location $(l_1,l_2)$ is labeled
$\mathbf{x}$ for simplicity. We construct the PDF of the WFCs in
this wavelet mode, $f(k)$, by looping over the other two indices
$(l_1,l_2)$, and express this measurement as a convolution of the
PDF of two components. The Gaussian contribution $G(\mathbf{x}) =
K(\mathbf{x}) - N(\mathbf{x})$ has a Gaussian PDF by assumption. We
label the PDF of the non-Gaussian contribution as $\Theta(n)$, and
write
\begin{equation}\label{fd}
    f(k)=\frac{1}{\sqrt{2\pi}}\int\Theta(n)\exp(-\frac{1}{2}(k-n)^2){\rm
    d}n
\end{equation}
This description also assumes that all the measurements of
$\mathbf{x}$ are statistically independent. Thanks to Bayes's
theorem, we can then calculate the conditional probability
$P(N|K)=P(K|N)P(N)/P(K)$. The posterior conditional expectation
value is written as
\begin{eqnarray}\label{s}
    \langle N|K=k \rangle & = & \frac{1}{\sqrt{2\pi}f(k)}\int{\exp[-\frac{1}{2}(n-k)^2]\Theta(n)n{\rm d}n} \nonumber \\
    &=&K+\frac{1}{\sqrt{2\pi}f(k)}\partial_k\int{\exp[-\frac{1}{2}(n-k)^2]\Theta(n){\rm d}n} \nonumber \\
    &=&K+(\ln f)'(k)
\end{eqnarray}
The expectation value of the Gaussian distribution is simply
recovered from  Eq.(\ref{dabw}). In other words, knowing $f(k)$
allows one to solve for both $G$ and $N$ given $K$.

When measuring the PDF of a given wavelet mode $(j_1,j_2)$ in
practice, we encounter the difficulty of poorly determined PDFs, and
thus even worse are the filter functions. This condition is rooted
in the scarceness of the WFCs (especially for larger scale wavelet
modes, containing less WFCs) and in the instability of numerical
differentiation. In order to resolve this issue, we construct $f(k)$
for each wavelet mode by looping over the spacial indices
$(l_1,l_2)$, and also over all the realizations. We further improve
the resolution of the PDF of each mode with a two parameter
analytical function:
\begin{equation}\label{fPDF}
    f_{\rm PDF}(x)=\frac{1}{\sqrt{\pi}s^{1-\alpha s^2}}
    \frac{\Gamma(\frac{1}{2}\alpha s^2)}{\Gamma(\frac{1}{2}\alpha s^2-\frac{1}{2})}
    (s^2-x^2)^{-\frac{\alpha s^2}{2}}
\end{equation}
which follow the PDF curves (see Fig.\ref{f.pdfs}), where the
$\Gamma$'s are the usual Gamma functions. The two parameters
$\alpha$ and $s$ are actually extracted via their relationship to
the second moment $m_2$ and the fourth moment $m_4$ of the PDF:
\begin{equation}\label{get_alpha}
    \alpha=\frac{5m_4-9m_2^2}{2m_2 m_4}
\end{equation}
and
\begin{equation}\label{get_s}
    s=\sqrt{\left|\frac{2m_2 m_4}{m_4-3m_2^2}\right|}.
\end{equation}

Having measured the expectation value of both components, we can now
loop back over all spatial indices $(l_1,l_2)$ and separate each
coefficient into the two components, by applying the following
(non-linear) Wiener filters:
 \begin{equation}\label{alpha}
    w_{\rm g}(\mathbf{x})=G/K = -\frac{(\ln f)'(k)}{k}
\end{equation}
\begin{equation}\label{beta}
    w_{\rm ng}(\mathbf{x})=N/K = 1+\frac{(\ln f)'(k)}{k}
\end{equation}
which are functions of $K$ only, and, we recall, apply solely to the
current wavelet mode $(j_1,j_2)$.

With the above parametrization in terms of $(\alpha, s)$, the
Gaussian filter function can be expressed by combining
Eq.(\ref{alpha}) with Eq.(\ref{fPDF}):
\begin{equation}
    w_{\rm g}(x)=\left(1+\frac{x^2}{s^2} \right)^{-1}.
\end{equation}
Note that the final filter function depends only on $s$, which
characterizes the extent of the departure from a Gaussian PDF, i.e.,
the greater the $s$, the smaller departure from Gaussian.

In the limit of a pure Gaussian PDF
\begin{equation}
    f_{\rm PDF}^{\rm G}=\mathcal{N}(0,\sigma^2)=\frac{1}{\sqrt{2\pi}\sigma}
    e^{-\frac{x^2}{2\sigma^2}},
\end{equation}
the second moment and the fourth moment are given by $m_2^{\rm
G}=\sigma^2$ and $m_4^{\rm G}=3\sigma^4$. From
Eq.(\ref{get_alpha},\ref{get_s}) we have $\alpha=\sigma^{-2}$ and
$s\rightarrow+\infty$, which is understandable. In this case, from
Eq.(\ref{alpha}) the filter function $w_{\rm g}^{\rm G}$ reduces to
a constant unity function $w_{\rm g}^{\rm G}(x)=1$. This is to be
expected since for those Gaussian or nearly Gaussian distributed
PDFs, we do not need to filter at all.

In Fig.\ref{f.pdfs}, we select four wavelet modes, characterizing
four different scales of the $\kappa$ fields, and plot the PDF and
fitted PDF for each of their WFCs. In order to visualize better the
non-Gaussianities of the PDF, we also plot for each panel a Gaussian
PDF which fits the central region of the fitted PDF Eq.(\ref{fPDF}).
As indicated by the dilations $(j_1,j_2)$ in each of the panel's
title, the corresponding scale increases when one looks from the
upper left panel to the bottom right panel. The corresponding
multipole moments for those four chosen $(j_1,j_2)$'s are roughly
$\ell\simeq9.1\times 10^3$, $2.3\times 10^3$, $5.7\times 10^2$ and
$2.8\times 10^2$, respectively. One can also see that, as the scale
goes up (or $\ell$ goes down), the parameter $s$ grows larger, and
the PDF (black solid line) of that wavelet mode becomes more
Gaussian, and closer to the Gaussian distribution (blue dashed
line).

Note that the wavelet non-linear Wiener filter is a parameter-free
method. In each wavelet mode, the filter is determined only by the
1-point PDF of all the WFCs. The method is stable in that, if we
make little changes in data in real space, then all the WFC's PDFs
will have little change, but will not have much effect on the final
fit, thus will not change the non-Gaussian decomposition. In
contrast, if one adds in a lot of non-Gaussian features on certain
scales, then the PDF of the WFCs, especially on that scale, will be
more non-Gaussian. This results in a stronger filtering and those
structures will be filtered out.

\begin{figure}
    \includegraphics[width=.5\textwidth]{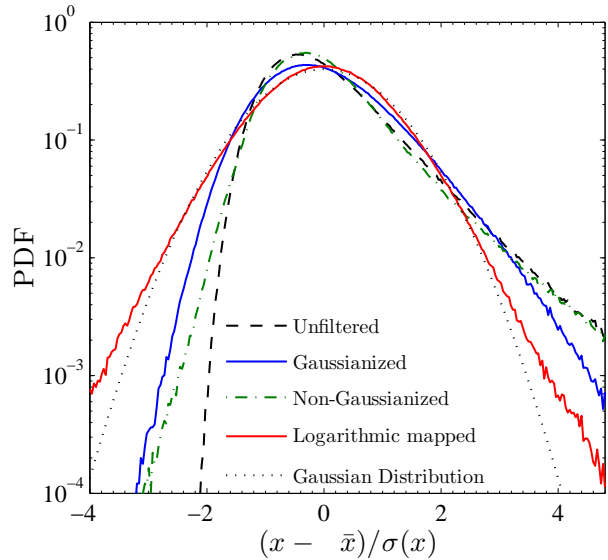}
    \caption{Probability distribution function of $\kappa$,
    $\kg$, $\kng$ and $\kln$ field. For comparison, A standard Gaussian
    distribution's PDF is also shown with a dotted line.}
    \label{f.pdf}
\end{figure}

We finally repeat this process for all wavelet modes, constructing
the filters $w_{\rm g}$ and $w_{\rm ng}$ and separate each $\kappa$
map into a Gaussian and non-Gaussian part. As an example, one
decomposition among those 185 realizations is shown in the middle
and right panel of Fig.\ref{f.maps}.

\subsection{Logarithmic Mapping of $\kappa$ Fields}\label{s.log}

In order to quantify the performance of our method, we wish to
compare the recovery of Fisher information in our Gaussianized
fields with that of a competing method. We follow the prescription
of \citep{2011ApJ...729L..11S}, and ``log-map'' each of our $\kappa$
fields
\begin{equation}\label{log_map}
    \kappa_{\rm ln} \equiv \kappa_0 \ln \left(1+\frac{\kappa}{\kappa_0}
    \right),
\end{equation}
where $\kappa_0$ is defined so as to keep the argument of logarithm
positive, i.e. $\kappa_0 < \min(\kappa)$, with $\min(\kappa)$ being
the minimum value of the $\kappa$ field. This transformation is
designed such that the non-linear peaks, who show strong convergence
values, are attenuated. As a result, higher-order statistics become
less important, which translates in a decrease of the non-Gaussian
contribution. We can characterize the degree of alteration of the
field by defining
 $r\equiv \kappa_0/|\min(\kappa)|$, where $1<r<\infty$. The smaller
the $r$, the more the log-mapping alters the field. In order to draw
general conclusions, we sample the $r$ space and try different
values in the following sections. These log-transforms are also
applied onto wavelet filtered $\kappa$ fields:
\begin{equation}
    \kappa_{\rm ln+g} \equiv \kappa_0 \ln \left(1+\frac{\kg}{\kappa_0}
    \right),
\end{equation}
The results of both methods and of their combination are discussed
in the next section.

\begin{figure}
    \includegraphics[width=.5\textwidth]{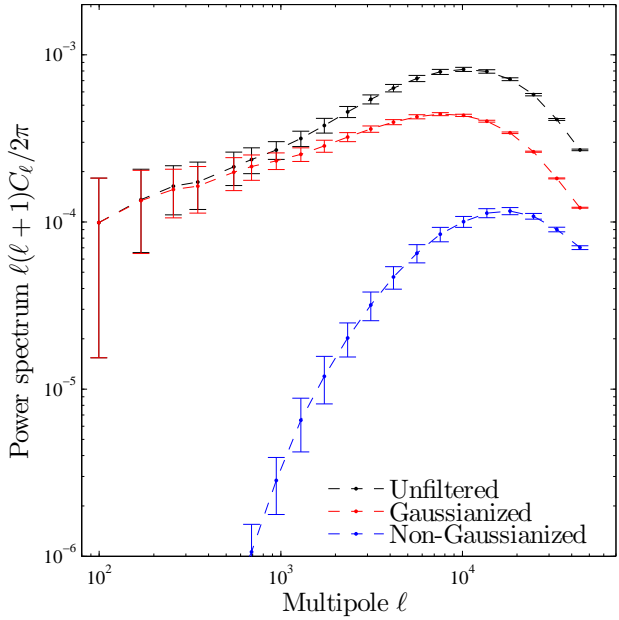}
    \caption{Angular power spectra for $\kappa$, $\kg$, and
    $\kng$ fields, respectively.}
    \label{f.pow}
\end{figure}

\begin{figure*}
    \centerline{
    \psfig{figure=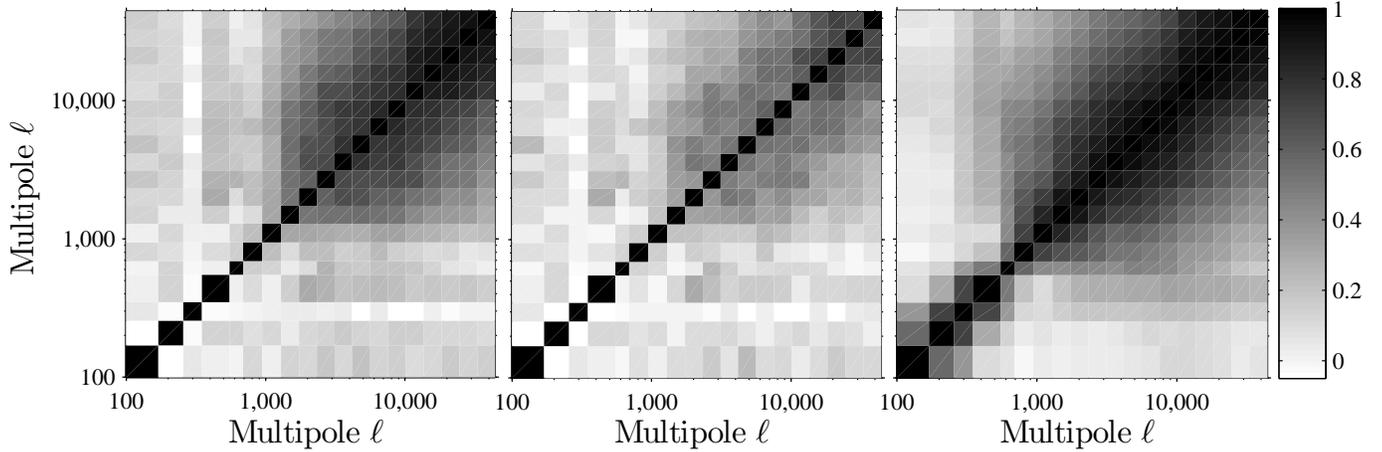,width=8.2truein,height=2.9truein}
    \hskip 0.0in}
    \caption{Cross-correlation coefficient
    matrices as found from 185 angular power spectra of $\kappa$ (left panel),
    $\kg$ (middle panel), and $\kng$ (right panel) fields respectively. The
    squares in black on the diagonal line indicate perfect correlation.}
    \label{f.matrix}
\end{figure*}

\section{Analysis}\label{s.analysis}

In Fig.\ref{f.maps}, we can see by eye that the peaks in the
unfiltered $\kappa$ maps are mostly transferred into $\kng$, leaving
$\kg$ with much less structure. We also plot the real-space 1-point
probability distribution function in Fig.\ref{f.pdf}. The result
shows that the original convergence map has a very large skewness.
We observe that the logarithmic mapping method have the best effect
in recovering a Gaussian PDF, while the non-linear Wiener filtering
mildly removes some of the skewness. This shows that the wavelet
techniques is optimized for restoring Gaussian features in the
2-point function, whereas logarithmic mapping is more effective on
the PDF. Both of these statistical estimators are useful in data
analysis, as they are sensitive to different systematics and probing
the cosmology in slightly different ways.

\subsection{Angular Power Spectra}
We calculate for each map their dimensionless angular power spectra
$\mathcal {C}(\ell)$, which is defined as
\begin{equation}\label{C_ell}
    \mathcal {C}(\ell)\equiv \frac{\ell(\ell+1)P(\ell)}{2\pi},
\end{equation}
where $\ell$ is the multipole characterizes the scale in the
two-dimensional $\kappa$-fields. We construct 20 $\ell$-bins and for
each wavenumber $|\vec{\ell}|$, the power spectrum $P(\ell)$ is
estimated by averaging all the Fourier modes within the bin, while
$\ell \equiv \langle\vec{\ell}\rangle_{\rm bin}$ is determined by
averaging all $|\vec{\ell}|$ that fall into the bin. In
Fig.\ref{f.pow} we plot the mean angular power spectra and error
bars of $\kappa$, $\kg$, and $\kng$ fields respectively. We can see
that on linear scales, where $\ell\lesssim 10^3$, the Gaussianized
power spectrum is nearly unchanged. On non-linear scales, however,
it drops by a factor of two, while the error bars are reduced.
Hence, in the presence of moderate noise, such a lowering of the
power spectrum shall not affect significantly the detectability.

\subsection{Cumulative Information}
We measure the covariance matrix of the $\mathcal {C}
(\ell)$, which captures the correlation between the variance, or the
error bars, of the power spectra at different scales $\ell$. If the
measurements were completely uncorrelated, the diagonal of the
covariance matrix would be the variance at each value of $\ell$, and
all the off-diagonal entries would be zero. Mathematically, the
covariance matrix $C$ is defined as
\begin{equation}
    C(\ell,\ell')\equiv \frac{1}{N-1}\sum_{i=1}^N
    [\mathcal {C}_i(\ell)-\langle \mathcal {C}(\ell)\rangle]
    [\mathcal {C}_i(\ell')-\langle \mathcal {C}(\ell')\rangle],
\end{equation}
where $N$ is the number of realizations and $\langle \mathcal
{C}(\ell)\rangle$ is the mean angular power spectrum over all
realizations. Note that the covariance matrix $C(\ell,\ell')$ is not
to be confused with angular power spectrum $\mathcal {C}(\ell)$.

The cross-correlation coefficient matrix, or for short the
correlation matrix, is a normalized version of the covariance
matrix, where each value is divided by the square root of the
diagonal values as follows:
\begin{equation}\label{rho}
    \rho(\ell,\ell')= \frac{C(\ell,\ell')}{\sqrt{C(\ell,\ell)C(\ell',\ell')}}.
\end{equation}
The three correlation matrices for our unfiltered, Gaussianized and
non-Gaussianized $\kappa$ fields are shown in the three panels in
Fig.\ref{f.matrix}. As expected, the original correlation matrix
shows cross-correlation greater than $60\%$ and up to $80\%$ where
$\ell\gtrsim 10^3$, while on larger scales, which is still in the
linear regime, the matrix is diagonal. In the Gaussianized
correlation matrix, the correlated region is much smaller, so the
correlations between different Fourier modes are suppressed to some
extent. In contrast, in the non-Gaussianized matrix, the
cross-correlation is even higher and even spread into linear region.

\begin{figure}
    \includegraphics[width=.5\textwidth]{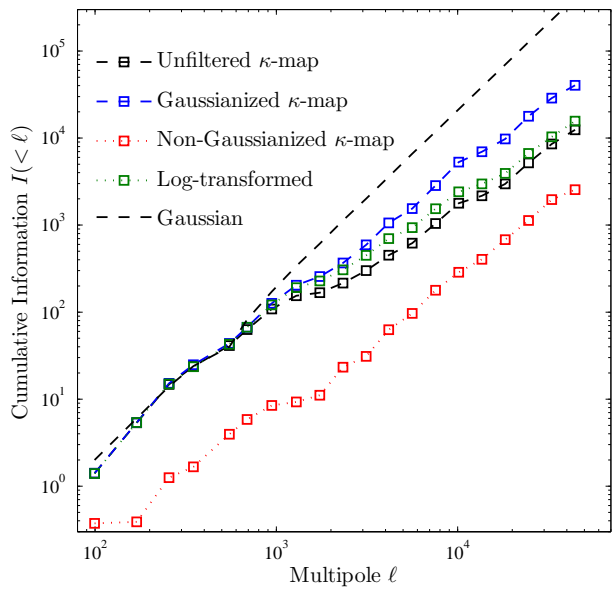}
    \caption{Cumulative information contained in the angular power
    spectra of the $\kappa$, $\kg$, $\kng$ and $\kln$ fields
    respectively. The Gaussian (linear) information is shown with the dashed line.}
    \label{f.info}
\end{figure}

The cumulative, or Fisher, information, is a measurement of the
number of independent Fourier modes presented in a field up to a
given scale, here $\ell_{n}$. It  is obtained as follows: for a
given wavenumber $\ell_{n}$, we select the subsection of the
covariance matrix up to that scale, then we invert this sub-matrix
and sum over all its elements. The results are meaningful when using
a normalized covariance matrix, which is defined as
\begin{equation}\label{Cnorm}
    C_{\rm norm}(\ell,\ell')=\frac{C(\ell,\ell')}{\langle \mathcal {C}(\ell)\rangle \langle
    \mathcal {C}(\ell')\rangle},
\end{equation}
and the function of cumulative information is
\begin{equation}\label{cumuinfo}
    I(<\ell_n)=\sum_{i,j=1}^n C_{\rm norm}^{-1}(\ell_i,\ell_j).
\end{equation}
We plot the cumulative information contained in the unfiltered,
Gaussianized and non-Gaussianized angular power spectra respectively
in Fig.\ref{f.info}. We also over-plot the information curve
obtained from $\kln$ with $r=1.5$ which seems to have the best
result among log transforms, and the theoretical Gaussian
information. In analogy with the Fisher information measured in the
density field
\cite{2005MNRAS.360L..82R,2006MNRAS.371.1205R,2011ApJ...728...35Z},
$\kappa$-fields' information also deviates from the Gaussian
prediction towards $\ell = 800$. However, in the weak lensing case,
the departure from Gaussian predictions is somehow attenuated, an
effect of the angular projection across multiple scales.

Note that because we are using a finite number of simulations to
measure several bins of data, even for a Gaussian random field, the
inverse of the covariance matrix is biased
\citep{2007A&A...464..399H}. The biases are bin-dependent and are up
to $10\%$, and would be applied to all the information curves, but
we are interested only in the difference between those curves. For
this reason we do not include this bias in all calculations.

We also note that our unfiltered Fisher information is somewhat
lower than that presented by \cite{2009arXiv0905.0501D}, but this
apparent discrepancy is caused by a different choice of galaxy
window function. \cite{2009arXiv0905.0501D} opted for a series of
top hat  windows, extending from $z\sim1.0$ to $z\sim3.0$, while we
include galaxy counts all the way down to $z=0$.  This effectively
enhances the amount of non-linearities in our fields, thus reducing
the Fisher information.

After performing the wavelet non-linear Wiener filtering and
logarithmic mapping method, both the $\kg$ and $\kln$ fields
recovered information, however, the wavelet technique restores
nearly 4 times more information  than the logarithmic mapping for
$\ell > 10000$. This performance gap gets even larger for $r=3.0$,
$r=2.0$ or $r=1.05$. We should also mention that wavelet method has
the advantage to be a parameter-free method, in the sense that
degrading the resolution by factors of two does not affect the
WFC¡¯s PDF estimation of other scales. For the logarithmic method,
however, one needs to re-optimize the value of $r$ for different
resolution, which affects the Fisher information at all scales. For
example, there are likely coarser resolutions where the log
transform would perform better (and worse) than the resolution shown
here.

We also investigated whether these two methods could be combined
together such that the total information recover exceeds that of the
two methods separately and found that these two methods should not
be used in conjunction. When we perform a logarithmic mapping after
wavelet filtering, the recovered Fisher information is lower than
for the wavelets alone. This is likely caused by the fact that a
logarithmic mapping applied on a nearly Gaussian field transforms
the field into something else, even less Gaussian. Because the
wavelet transform is resource consuming, we have not tried the other
way around, i.e. performing the logarithmic mapping first, which
would involve a filtering for each sampling of $\kappa_0$. However,
we have strong reasons to believe that the gain would also be
minimal, if not worse than the wavelet on its own. As discussed
before, we observe that the PDFs of the log-transformed fields, in
both real and wavelet spaces, are nearly Gaussian. This means that
the wavelet filtering would have a very small impact, as most of the
contribution would directly fall in the Gaussian filter. As a
result, the information that is recovered by the wavelet non-linear
Wiener filter, but {\it not} by the log-transform, will not reappear
with an additional Wiener filtering. However, other Gaussianization
methods, especially those that attempt to go back in time -- and
thus only leave some non-Gaussian features -- are likely to combine
much better with the wavelet non-linear Wiener filters. For example,
we are in the process of testing whether wavelet filtering can be
used in conjunction with a density reconstruction algorithm to
optimize the Fisher information.

In addition, it has been shown that the logarithmic mapping has very
little gain in Fisher information when one includes realistic levels
of Gaussian noise \citep{2011arXiv1109.5639S}, which are inherent to
weak lensing observations. Although we have not yet tested the
performance of wavelet filtering in such a noisy environment, it has
for sure the advantage of being able to extract a non-Gaussian
component, which still contains most of the collapsed structures,
and give us a novel handle in the data analysis.

The extension of this work, which we postpone to a future paper, is
to propagate the uncertainty of the measured angular power spectrum
onto that of cosmological parameters, following the Fisher formalism
summarized in \cite{2006astro.ph..9591A}, and to compare the
constraining performances of the covariance matrices from the
unfiltered and from the Gaussianized fields to the simple Gaussian
analysis. Quantifying the difference in uncertainty on the dark
energy figure-of-merit is the final objective, and a significant
improvement would lead to a enhanced yet robust precision on
$\omega$.

\section{Conclusion}\label{s.conclusion}
We ran a series of $N$-body simulations to generate weak lensing
convergence maps. After analyzing their angular power spectra, as
expected, we find that their cross correlation matrix is highly
correlated on small scales ($\ell\gtrsim 10^3$), since these scales
are in non-linear regime. This non-linear clustering makes the
Fisher information contained in the angular power spectra less than
Gaussian information. The lost information can be obtained by higher
statistics like bi-spectrum (three point correlation function),
tri-spectrum etc., but to avoid dealing with tedious higher order
calculations, we use two mathematical tools to Gaussianize $\kappa$
fields, namely a logarithmic mapping and non-linear wavelet Wiener
filters.

The method of wavelet non-linear Wiener filtering and logarithmic
mapping can both increase the Fisher information in the angular
power spectra, but wavelet method has better results with a gain of
a factor 4 at $\ell = 10 000$. However, one needs to keep in mind
that with the forecasted level of shot noise of future survey like
EUCLID \citep{2010ASPC..430..266B} and JDEM
\citep{2010arXiv1008.4936G}, the relative importance of
non-Gaussianities will be reduced significantly
\citep{2009arXiv0905.0501D}. We also find that these two methods are
not complementary, and that their combination reduces the
information recovery compared to a sole Gaussianization of the
fields.

\section*{Acknowledgements}
We are very grateful to the anonymous referee for many valuable
comments that greatly improved the paper. This work was supported by
the National Science Foundation of China (Grants No. 11173006), the
Ministry of Science and Technology National Basic Science program
(project 973) under grant No. 2012CB821804, and the Fundamental
Research Funds for the Central Universities. Computations were
performed on the TCS supercomputer at the SciNet HPC Consortium.
SciNet is funded by: the Canada Foundation for Innovation under the
auspices of Compute Canada; the Government of Ontario; Ontario
Research Fund - Research Excellence; and the University of Toronto.
UP and JHD would like to acknowledge NSERC for their financial
support.

\bibliographystyle{hapj}
\bibliography{msk_mn}

\begin{thebibliography}{60}
\expandafter\ifx\csname natexlab\endcsname\relax\def\natexlab#1{#1}\fi

\bibitem[{{Albrecht} {et~al.}(2006){Albrecht}, {Bernstein}, {Cahn}, {Freedman},
  {Hewitt}, {Hu}, {Huth}, {Kamionkowski}, {Kolb}, {Knox}, {Mather}, {Staggs},
  \& {Suntzeff}}]{2006astro.ph..9591A}
{Albrecht}, A. {et~al.} 2006, ArXiv Astrophysics e-prints,
  arXiv:astro-ph/0609591

\bibitem[{{Bartelmann} \& {Schneider}(1999)}]{1999A&A...345...17B}
{Bartelmann}, M., \& {Schneider}, P. 1999, \aap, 345, 17,
  arXiv:astro-ph/9902152

\bibitem[{{Bartelmann} \& {Schneider}(2001)}]{2001PhR...340..291B}
------. 2001, \physrep, 340, 291, arXiv:astro-ph/9912508

\bibitem[{{Beaulieu} {et~al.}(2010){Beaulieu}, {Bennett}, {Batista}, {Cassan},
  {Kubas}, {Fouqu{\'e}}, {Kerrins}, {Mao}, {Miralda-Escud{\'e}}, {Wambsganss},
  {Gaudi}, {Gould}, \& {Dong}}]{2010ASPC..430..266B}
{Beaulieu}, J.~P. {et~al.} 2010, in Astronomical Society of the Pacific
  Conference Series, Vol. 430, Astronomical Society of the Pacific Conference
  Series, ed. {V.~Coud{\'e} Du Foresto, D.~M.~Gelino, \& I.~Ribas}, 266--+,
  1001.3349

\bibitem[{{Benjamin} {et~al.}(2007){Benjamin}, {Heymans}, {Semboloni}, {van
  Waerbeke}, {Hoekstra}, {Erben}, {Gladders}, {Hetterscheidt}, {Mellier}, \&
  {Yee}}]{2007MNRAS.381..702B}
{Benjamin}, J. {et~al.} 2007, \mnras, 381, 702, arXiv:astro-ph/0703570

\bibitem[{Daubechies(1992)}]{Daubechies1992}
Daubechies, I. 1992, Ten Lectures on Wavelets (C B M S - N S F Regional
  Conference Series in Applied Mathematics) ({Soc for Industrial \& Applied
  Math})

\bibitem[{{Dewdney} {et~al.}(2009){Dewdney}, {Hall}, {Schilizzi}, \&
  {Lazio}}]{2009IEEEP..97.1482D}
{Dewdney}, P.~E., {Hall}, P.~J., {Schilizzi}, R.~T., \& {Lazio}, T.~J.~L.~W.
  2009, IEEE Proceedings, 97, 1482

\bibitem[{{Dor{\'e}} {et~al.}(2009){Dor{\'e}}, {Lu}, \&
  {Pen}}]{2009arXiv0905.0501D}
{Dor{\'e}}, O., {Lu}, T., \& {Pen}, U. 2009, ArXiv e-prints, 0905.0501

\bibitem[{{Eisenstein} {et~al.}(2007){Eisenstein}, {Seo}, {Sirko}, \&
  {Spergel}}]{2007ApJ...664..675E}
{Eisenstein}, D.~J., {Seo}, H., {Sirko}, E., \& {Spergel}, D.~N. 2007, \apj,
  664, 675, arXiv:astro-ph/0604362

\bibitem[{Eisenstein {et~al.}(2005)}]{Eisenstein:2005su}
Eisenstein, D.~J., {et~al.} 2005, Astrophys. J., 633, 560, astro-ph/0501171

\bibitem[{Fang \& Thews(1998)}]{1998World.Sci..2A}
Fang, L.-Z.~e., \& Thews, R. L.~e. 1998, Wavelets in Physics (Singapur: World
  Scientific)

\bibitem[{{Feldman} {et~al.}(1994){Feldman}, {Kaiser}, \&
  {Peacock}}]{1994ApJ...426...23F}
{Feldman}, H.~A., {Kaiser}, N., \& {Peacock}, J.~A. 1994, \apj, 426, 23

\bibitem[{{Fisher}(1935)}]{Fisher1935}
{Fisher}, R.~A. 1935, Journal of the Royal Statistical Society, 98, 39

\bibitem[{{Gehrels}(2010)}]{2010arXiv1008.4936G}
{Gehrels}, N. 2010, ArXiv e-prints, 1008.4936

\bibitem[{{Goldberg} \& {Spergel}(2000)}]{2000ASPC..201..282G}
{Goldberg}, D.~M., \& {Spergel}, D.~N. 2000, in Astronomical Society of the
  Pacific Conference Series, Vol. 201, Cosmic Flows Workshop, ed. {S.~Courteau
  \& J.~Willick}, 282--+, arXiv:astro-ph/9909057

\bibitem[{{Harnois-D{\'e}raps} \& {Pen}(2011)}]{2011arXiv1109.5746H}
{Harnois-D{\'e}raps}, J., \& {Pen}, U.-L. 2011, ArXiv e-prints, 1109.5746

\bibitem[{{Hartlap} {et~al.}(2007){Hartlap}, {Simon}, \&
  {Schneider}}]{2007A&A...464..399H}
{Hartlap}, J., {Simon}, P., \& {Schneider}, P. 2007, \aap, 464, 399,
  arXiv:astro-ph/0608064

\bibitem[{{Hoekstra} {et~al.}(2006){Hoekstra}, {Mellier}, {van Waerbeke},
  {Semboloni}, {Fu}, {Hudson}, {Parker}, {Tereno}, \&
  {Benabed}}]{2006ApJ...647..116H}
{Hoekstra}, H. {et~al.} 2006, \apj, 647, 116, arXiv:astro-ph/0511089

\bibitem[{{Huterer}(2002)}]{2002PhRvD..65f3001H}
{Huterer}, D. 2002, \prd, 65, 063001, arXiv:astro-ph/0106399

\bibitem[{{Huterer}(2010)}]{2010GReGr..42.2177H}
------. 2010, General Relativity and Gravitation, 42, 2177, 1001.1758

\bibitem[{{Joachimi} {et~al.}(2011){Joachimi}, {Taylor}, \&
  {Kiessling}}]{2011arXiv1104.1399J}
{Joachimi}, B., {Taylor}, A.~N., \& {Kiessling}, A. 2011, ArXiv e-prints,
  1104.1399

\bibitem[{{Kaiser} \& {Squires}(1993)}]{1993ApJ...404..441K}
{Kaiser}, N., \& {Squires}, G. 1993, \apj, 404, 441

\bibitem[{{Loken} {et~al.}(2010){Loken}, {Gruner}, {Groer}, {Peltier}, {Bunn},
  {Craig}, {Henriques}, {Dempsey}, {Yu}, {Chen}, {Dursi}, {Chong}, {Northrup},
  {Pinto}, {Knecht}, \& {Van Zon}}]{Scinet}
{Loken}, C. {et~al.} 2010, Journal of Physics: Conference Series, 256

\bibitem[{{LSST Science Collaborations} {et~al.}(2009){LSST Science
  Collaborations}, {Abell}, {Allison}, {Anderson}, {Andrew}, {Angel}, {Armus},
  {Arnett}, {Asztalos}, {Axelrod}, \& et~al.}]{2009arXiv0912.0201L}
{LSST Science Collaborations} {et~al.} 2009, ArXiv e-prints, 0912.0201

\bibitem[{{Lu} {et~al.}(2010){Lu}, {Pen}, \& {Dor{\'e}}}]{2010PhRvD..81l3015L}
{Lu}, T., {Pen}, U., \& {Dor{\'e}}, O. 2010, \prd, 81, 123015, 0905.0499

\bibitem[{{Massey} {et~al.}(2010){Massey}, {Kitching}, \&
  {Richard}}]{2010RPPh...73h6901M}
{Massey}, R., {Kitching}, T., \& {Richard}, J. 2010, Reports on Progress in
  Physics, 73, 086901, 1001.1739

\bibitem[{{Mellier}(1999)}]{1999toc..conf..211M}
{Mellier}, Y. 1999, in NATO ASIC Proc. 541: Theoretical and Observational
  Cosmology, 211--+, arXiv:astro-ph/9901116

\bibitem[{{Merz} {et~al.}(2005){Merz}, {Pen}, \& {Trac}}]{2005NewA...10..393M}
{Merz}, H., {Pen}, U.-L., \& {Trac}, H. 2005, New Astronomy, 10, 393,
  arXiv:astro-ph/0402443

\bibitem[{{Neyrinck} \& {Szapudi}(2007)}]{2007MNRAS.375L..51N}
{Neyrinck}, M.~C., \& {Szapudi}, I. 2007, \mnras, 375, L51,
  arXiv:astro-ph/0610211

\bibitem[{{Neyrinck} {et~al.}(2006){Neyrinck}, {Szapudi}, \&
  {Rimes}}]{2006MNRAS.370L..66N}
{Neyrinck}, M.~C., {Szapudi}, I., \& {Rimes}, C.~D. 2006, \mnras, 370, L66,
  arXiv:astro-ph/0604282

\bibitem[{{Neyrinck} {et~al.}(2009){Neyrinck}, {Szapudi}, \&
  {Szalay}}]{2009ApJ...698L..90N}
{Neyrinck}, M.~C., {Szapudi}, I., \& {Szalay}, A.~S. 2009, \apjl, 698, L90,
  0903.4693

\bibitem[{{Neyrinck} {et~al.}(2011){Neyrinck}, {Szapudi}, \&
  {Szalay}}]{2011ApJ...731..116N}
------. 2011, \apj, 731, 116, 1009.5680

\bibitem[{{Ngan} {et~al.}(2011){Ngan}, {Harnois-D{\'e}raps}, {Pen}, {McDonald},
  \& {MacDonald}}]{2011arXiv1106.5548N}
{Ngan}, W.-H.~W., {Harnois-D{\'e}raps}, J., {Pen}, U.-L., {McDonald}, P., \&
  {MacDonald}, I. 2011, ArXiv e-prints, 1106.5548

\bibitem[{{Noh} {et~al.}(2009){Noh}, {White}, \&
  {Padmanabhan}}]{2009PhRvD..80l3501N}
{Noh}, Y., {White}, M., \& {Padmanabhan}, N. 2009, \prd, 80, 123501, 0909.1802

\bibitem[{{Oguri} {et~al.}(2008){Oguri}, {Inada}, {Strauss}, {Kochanek},
  {Richards}, {Schneider}, {Becker}, \& {Fukugita}}]{2008AJ....135..512O}
{Oguri}, M., {Inada}, N., {Strauss}, M.~A., {Kochanek}, C.~S., {Richards},
  G.~T., {Schneider}, D.~P., {Becker}, R.~H., \& {Fukugita}. 2008, \aj, 135,
  512, 0708.0825

\bibitem[{{Padmanabhan} {et~al.}(2009){Padmanabhan}, {White}, \&
  {Cohn}}]{2009PhRvD..79f3523P}
{Padmanabhan}, N., {White}, M., \& {Cohn}, J.~D. 2009, \prd, 79, 063523,
  0812.2905

\bibitem[{{Pen}(1999)}]{1999RSPTA.357.2561P}
{Pen}, U.-L. 1999, Royal Society of London Philosophical Transactions Series A,
  357, 2561, arXiv:astro-ph/9904170

\bibitem[{{Percival} {et~al.}(2007){Percival}, {Nichol}, {Eisenstein},
  {Weinberg}, {Fukugita}, {Pope}, {Schneider}, {Szalay}, {Vogeley}, {Zehavi},
  {Bahcall}, {Brinkmann}, {Connolly}, {Loveday}, \&
  {Meiksin}}]{2007ApJ...657...51P}
{Percival}, W.~J. {et~al.} 2007, \apj, 657, 51, arXiv:astro-ph/0608635

\bibitem[{{Perlmutter} {et~al.}(1999){Perlmutter}, {Aldering}, {Goldhaber},
  {Knop}, {Nugent}, {Castro}, {Deustua}, \& {The Supernova Cosmology
  Project}}]{1999ApJ...517..565P}
{Perlmutter}, S., {Aldering}, G., {Goldhaber}, G., {Knop}, R.~A., {Nugent}, P.,
  {Castro}, P.~G., {Deustua}, S.~{Couch}, W.~J., \& {The Supernova Cosmology
  Project}. 1999, \apj, 517, 565

\bibitem[{{Peterson} {et~al.}(2006){Peterson}, {Bandura}, \&
  {Pen}}]{2006astro.ph..6104P}
{Peterson}, J.~B., {Bandura}, K., \& {Pen}, U.~L. 2006, ArXiv Astrophysics
  e-prints, arXiv:astro-ph/0606104

\bibitem[{{Press} {et~al.}(1992){Press}, {Teukolshy}, {Vetterling}, \&
  {Flannery}}]{nr}
{Press}, W., {Teukolshy}, S., {Vetterling}, W., \& {Flannery}, B. 1992,
  {Numerical Recipes in FORTRAN. The second edition} (Cambridge University
  Press, 1992)

\bibitem[{{Refregier}(2003)}]{2003ARA&A..41..645R}
{Refregier}, A. 2003, \araa, 41, 645, arXiv:astro-ph/0307212

\bibitem[{{Rimes} \& {Hamilton}(2005)}]{2005MNRAS.360L..82R}
{Rimes}, C.~D., \& {Hamilton}, A.~J.~S. 2005, \mnras, 360, L82,
  arXiv:astro-ph/0502081

\bibitem[{{Rimes} \& {Hamilton}(2006)}]{2006MNRAS.371.1205R}
------. 2006, \mnras, 371, 1205, arXiv:astro-ph/0511418

\bibitem[{{Schilizzi}(2007)}]{2007HiA....14..539S}
{Schilizzi}, R.~T. 2007, Highlights of Astronomy, 14, 539

\bibitem[{{Schlegel} {et~al.}(2009){Schlegel}, {White}, \&
  {Eisenstein}}]{2009astro2010S.314S}
{Schlegel}, D., {White}, M., \& {Eisenstein}, D. 2009, in Astronomy, Vol. 2010,
  astro2010: The Astronomy and Astrophysics Decadal Survey, 314--+, 0902.4680

\bibitem[{{Seljak}(1998)}]{1998ApJ...506...64S}
{Seljak}, U. 1998, \apj, 506, 64

\bibitem[{{Seo} {et~al.}(2011{\natexlab{a}}){Seo}, {Sato}, {Dodelson}, {Jain},
  \& {Takada}}]{2011ApJ...729L..11S}
{Seo}, H.-J., {Sato}, M., {Dodelson}, S., {Jain}, B., \& {Takada}, M.
  2011{\natexlab{a}}, \apjl, 729, L11+, 1008.0349

\bibitem[{{Seo} {et~al.}(2011{\natexlab{b}}){Seo}, {Sato}, {Takada}, \&
  {Dodelson}}]{2011arXiv1109.5639S}
{Seo}, H.-J., {Sato}, M., {Takada}, M., \& {Dodelson}, S. 2011{\natexlab{b}},
  ArXiv e-prints, 1109.5639

\bibitem[{{Takahashi} {et~al.}(2011){Takahashi}, {Yoshida}, {Takada},
  {Matsubara}, {Sugiyama}, {Kayo}, {Nishimichi}, {Saito}, \&
  {Taruya}}]{2011ApJ...726....7T}
{Takahashi}, R. {et~al.} 2011, \apj, 726, 7, 0912.1381

\bibitem[{{Tegmark} {et~al.}(2006){Tegmark}, {Eisenstein}, {Strauss},
  {Weinberg}, {Blanton}, {Frieman}, {Fukugita}, {Gunn}, \&
  {Hamilton}}]{2006PhRvD..74l3507T}
{Tegmark}, M. {et~al.} 2006, \prd, 74, 123507, arXiv:astro-ph/0608632

\bibitem[{{Tegmark} {et~al.}(1997){Tegmark}, {Taylor}, \&
  {Heavens}}]{1997ApJ...480...22T}
{Tegmark}, M., {Taylor}, A.~N., \& {Heavens}, A.~F. 1997, \apj, 480, 22,
  arXiv:astro-ph/9603021

\bibitem[{{Vafaei} {et~al.}(2010){Vafaei}, {Lu}, {van Waerbeke}, {Semboloni},
  {Heymans}, \& {Pen}}]{2010APh....32..340V}
{Vafaei}, S., {Lu}, T., {van Waerbeke}, L., {Semboloni}, E., {Heymans}, C., \&
  {Pen}, U.-L. 2010, Astroparticle Physics, 32, 340, 0905.3726

\bibitem[{{Van Waerbeke} {et~al.}(2001){Van Waerbeke}, {Mellier}, {Radovich},
  {Bertin}, {Dantel-Fort}, {McCracken}, {Le F{\` e}vre}, {Foucaud},
  {Cuillandre}, {Erben}, {Jain}, {Schneider}, {Bernardeau}, \&
  {Fort}}]{2001A&A...374..757V}
{Van Waerbeke}, L. {et~al.} 2001, \aap, 374, 757

\bibitem[{{Vogeley} \& {Szalay}(1996)}]{1996ApJ...465...34V}
{Vogeley}, M.~S., \& {Szalay}, A.~S. 1996, \apj, 465, 34,
  arXiv:astro-ph/9601185

\bibitem[{{Voit}(2005)}]{2005RvMP...77..207V}
{Voit}, G.~M. 2005, Reviews of Modern Physics, 77, 207, arXiv:astro-ph/0410173

\bibitem[{{Weinberg}(1992)}]{1992MNRAS.254..315W}
{Weinberg}, D.~H. 1992, \mnras, 254, 315

\bibitem[{{Yu} {et~al.}(2011){Yu}, {Zhang}, {Lin}, {Cui}, \&
  {Fry}}]{2011arXiv1103.2858Y}
{Yu}, Y., {Zhang}, P., {Lin}, W., {Cui}, W., \& {Fry}, J.~N. 2011, ArXiv
  e-prints, 1103.2858

\bibitem[{{Zhang} {et~al.}(2003){Zhang}, {Pen}, {Zhang}, \&
  {Dubinski}}]{2003ApJ...598..818Z}
{Zhang}, T., {Pen}, U., {Zhang}, P., \& {Dubinski}, J. 2003, \apj, 598, 818

\bibitem[{{Zhang} {et~al.}(2011){Zhang}, {Yu}, {Harnois-D{\'e}raps},
  {MacDonald}, \& {Pen}}]{2011ApJ...728...35Z}
{Zhang}, T.-J., {Yu}, H.-R., {Harnois-D{\'e}raps}, J., {MacDonald}, I., \&
  {Pen}, U.-L. 2011, \apj, 728, 35, 1008.3506

\end{thebibliography}

\label{lastpage}

\end{document}